\documentclass[aps,prd,groupedaddress,twocolumn,floatfix,nofootinbib,showpacs,showkeys]{revtex4-1}

\usepackage{graphicx}
\usepackage{dcolumn}
\usepackage{bm}
\usepackage{amsmath}
\usepackage{epstopdf}
\usepackage[flushleft]{threeparttable}
\usepackage{color}

\usepackage[thinlines]{easytable}
\usepackage{array}
\usepackage{booktabs}
\usepackage{multirow}

\def\mnras{MNRAS}
\def\apj{Astroph. J.}

\def\aap{Astron. Astroph.}

\begin{document}

\title{Universal relations for the Keplerian sequence of rotating neutron stars}

\author{R.~Riahi$^{1,2}$}
\email{r.riahi@ph.iut.ac.ir}
\author{S.~ Z.~Kalantari$^{1}$}
\email{zafar@cc.iut.ac.ir}
\author{J.~A.~Rueda$^{2,3,4}$}
\email{jorge.rueda@icra.it}
\affiliation{$^1$Department of Physics, Isfahan University of Technology, 84156-83111, Iran}
\affiliation{$^2$ICRANet, Piazza della Repubblica 10, I--65122 Pescara, Italy}
\affiliation{$^3$Dipartimento di Fisica and ICRA, Sapienza Universit\`a di Roma, P.le Aldo Moro 5, I--00185 Rome, Italy}
\affiliation{$^4$ICRANet-Rio, Centro Brasileiro de Pesquisas F\'isicas, Rua Dr. Xavier Sigaud 150, Rio de Janeiro, RJ, 22290--180, Brazil}

\date{\today}

\begin{abstract}
We investigate the Keplerian (mass-shedding) sequence of rotating neutron stars. Twelve different equations of state are used to describe the nuclear structure. We find four fitting relations which connect the rotating frequency, mass and radius of stars in the mass-shedding limit to the mass and radius of stars in the static sequence. We show the breakdown of approximate relation for the Keplerian frequency derived by Lattimer and Prakash [Science, 304, 536, (2004)] and then we present a new, EOS-independent and more accurate relation. This relation fits the Keplerian frequency of rotating neutron stars to about $2\%$ for a large range of the compactness $M_{S}/R_{S}$ of the reference non-rotating neutron star, namely the static star with the same central density as the rotating one. The performance of the fitting formula is close to $4\%$ for $M_{S}/R_{S}\leq 0.05~M_{\odot}$/km ($f_{K}\leq 350$~Hz). We present additional EOS-independent relations for the Keplerian sequence including relations for $M_{K}f_{K}$ and $R_{K}f_{K}$ in terms of $M_{S}f_{S}$ and $R_{S}f_{S}$, respectively, one of $M_K/R_K$ as a function of $f_{K}/f_{S}$ and $M_S/R_S$, and a relation between the $M_K$, $R_K$ and $f_K$. These new fitting relations are approximately EOS-independent with an error in the worst case of $8\%$. The universality of the Keplerian sequence properties presented here add to the set of other neutron star universal relations in the literature such as the $I$-Love-$Q$ relation, the gravitational binding energy and the energy, angular momentum and radius of the last circular orbit of a test-particle around rotating neutron stars. This set of universal, analytic formulas, facilitate the inclusion of general relativistic effects in the description of relativistic astrophysical systems involving fast rotating neutron stars.
%
\end{abstract}



\maketitle

\section{Introduction}

Neutron stars are the densest stars in the Universe and play a fundamental role in the understanding of a large number of relativistic physics and astrophysics issues, e.g. the behavior of matter at supra-nuclear densities, the mechanisms that trigger the most energetic cataclysmic events in the Universe (supernovae and gamma-ray bursts), the formation of heavy nuclei via r-process nucleosynthesis in the ejecta of neutron star binary mergers and the population of high-frequency (kHz) gravitational-waves sources. In addition, the observations of neutron stars in less energetic systems, e.g. pulsars in binary systems, the thermal X-ray emission from isolated neutron stars and of the quasi-periodic oscillations from accreting neutron stars, is important to obtain information about neutron stars radii, masses, ages, internal composition and temperatures. 

The relation between the structure of a neutron star (mass, radius, etc) and the microphysical input, namely the equation of state (EOS) of ultra-dense matter, is crucial for the understanding of all the above physical and astrophysical scenarios. In view of the absence of a complete knowledge of the nuclear EOS at supra-nuclear conditions, a variety of approaches has been used for modeling neutron stars and deriving their properties which depend on the selected EOS \cite{haensel2007neutron,0004-637X-722-1-33,Lattimer2011,2013A&A...551A..61Z,LATTIMER2016127}. These properties have been compared with the observational constraints that help to narrow down the physically plausible nuclear EOS (see e.g.~\cite{Ozel2006}).

Many attempts have been made to obtain approximate, EOS-independent relations between neutron star properties which become useful tools in astrophysical applications without suffering from the EOS indeterminacy. Those relations are usually EOS-independent to within $O(1\%)$ error and they have been called \emph{universal} relations (see e.g. \cite{yagi2017approximate} and references therein). 
For instance, Laarakkers and Poisson \cite{0004-637X-512-1-282} calculated the quadrupole moment of the mass distribution, $Q$, and concluded that for a fixed mass and EOS, the $Q$-dependence on the angular momentum $J$ can be fitted by a quadratic formula. In \cite{mnras/stt858} it was found a fitting relation between the parameter $Q M/J^{2}$, and the ratio of the circumferential radius of star to its Schwarzschild radius, $R/(2M)$, for rotating neutron stars within the Hartle-Thorne, slow-rotation approximation. In \cite{Lattimer536} it was introduced an EOS-independent relation for the maximum rotational frequency of stars in terms of mass and radius of non-rotating stars. In \cite{0004-637X-714-2-1234} it was connected the frequency and damping rate of the quadrupole f-mode to the mass and moment of inertia of non-rotating neutron stars. In \cite{PhysRevD.88.023009,Yagi365} there were derived relations, within the framework of the Hartle-Thorne approximation, which connect the moment of inertia $I$, the Love number and the quadrupole moment $Q$ of rotating neutron stars, called $I$-Love-$Q$ relations. Afterwards, in \cite{2014ApJ...781L...6D} it was demonstrated the breakdown of such relations in fast rotating stars, then \cite{PhysRevLett.112.121101} extended these relations to fast rotating by replacing the dimensional frequency with the dimensionless angular momentum, $j=c J/(G M^2_\odot)$. Finally, in \cite{PhysRevLett.112.201102} other dimensionless quantities such as $M\times f$ and $R\times f$ were used to extend these relations. In \cite{refId0} a pair of relations connecting the maximum and minimum masses of a rotating star of rotation frequency $f$, to the maximum mass of static configuration, were derived. Additional relations between the redshift (polar, forward and backward) and the minimum and maximum compactness of the star were also presented. It was there reproduced the formula suggested in \cite{2001ApJ...550..426L} and reformulated in \cite{2002A&A...396..917B}, with different coefficients. In \cite{PhysRevD.96.024046} formulas for the binding energy of static and rotating stars as a function of the gravitational mass and dimensionless angular momentum, $j$, where presented. Also, they provided a formula connect the maximum mass, i.e., along the secular instability limit, and $j$. Ref.~ \cite{0004-637X-861-2-141} studied the last circular orbit of a test particle around fast rotating neutron stars. They presented a pair of fitting relations that connect the radius and orbital frequency of this orbit to the rotation frequency $f$ and mass $M$ of the rotating neutron star.

The aim of this work is to investigate the Keplerian (mass-shedding) sequence of rotating neutron stars to search for EOS-independent relations useful for astrophysical applications. We present here universal relations connecting the mass and frequency of a configuration along the Keplerian sequence, to the mass and radius of the non-rotating (static) configuration with the same central density as the rotating star, and to the Keplerian frequency of a test-particle in an orbit of size equal to the static neutron star radius. Although we use $c=G=1$ geometric units in our calculations, we restore physical units to simplify the use of results in astrophysical situations. We use units $M_\odot$, km and Hz, respectively for the mass, radius and frequency.

\section{Keplerian (mass-shedding) sequence of rotating neutron stars}

The EOS, namely relation between energy density and pressure, is an essential requirement for describing the macroscopic properties of stars. The EOS is used as input to the Einstein field equations. Non-relativistic \cite{PhysRevC.85.035201} and relativistic models \cite{PhysRevC.90.055203} have been used to obtain stellar properties. In this work, we use 12 EOS with different theoretical features including microscopic calculations, relativistic mean-field and Skyrme mean-field to find the universal relations. All of our selected EOS used in the fits of the numerical data support a non-rotating neutron star with a maximum mass larger than $M_{S} \gtrsim 2.0~M_{\odot}$, consistent with the current observational constraints \cite{Demorest2010/10/28/print,Antoniadis1233232}. There are two EOS based on microscopic calculations, APR \cite{PhysRevC.58.1804} and BL \cite{2018A&A...609A.128B}, eight relativistic mean-field models ,BKA22, BKA24 \cite{PhysRevC.81.034323}, CMF \cite{PhysRevC.92.012801}, DDH$\delta$(with hyperons) \cite{Gaitanos200424,PhysRevC.90.045803}, DD-ME$\delta$ \cite{PhysRevC.84.054309}, G1 \cite{FURNSTAHL1997441}, GM1(with hyprons) \cite{PhysRevLett.67.2414,Douchin2001,Oertel2015} and TW99  \cite{Typel1999331} and two Skyrme mean-field models, SKa \cite{PhysRevC.92.055803} and SLy4 \cite{DOUCHIN2000107}, to find the fitting relations. In order to test the robustness of the universal relations and the obtained fits, we compute the relative error for three additional EOS, not used in the fitting procedure, and check their relative error lies with the error attached to the given fits. For this task we use a soft EOS, G2$^{*}$ \cite{PhysRevC.74.045806}, a moderately stiff EOS, GM1 (without hyperons)\cite{PhysRevLett.67.2414}, and a stiff EOS, NL3 \cite{PhysRevC.55.540}.

In order to solve Einstein field equations and investigate the structure and gravitational field of relativistic, axisymmetric, stationary and uniform rotating stars, many numerical methods have been developed since the 1970's (see e.g. \cite{1972ApJ...176..195W,1974ApJ...191..273B,PhysRevLett.62.3015,1989MNRAS.237..355K,1992LNP...410..305N,1995ApJ...444..306S}). Based on these methods a few publicly numerical codes have been developed \cite{Paschalidis2017} to solve them. We use the public LORENE library (https://lorene.obspm.fr) for the numerical solution. This code has being developed based on the multi-domain spectral method which had presented by Bonazzola et al. \cite{PhysRevD.58.104020,1993A&A...278..421B} and developed by Ansorg et al.~\cite{AnsorgM.2003}.

\begin{figure}
    \centering
    \includegraphics[width=\hsize,clip]{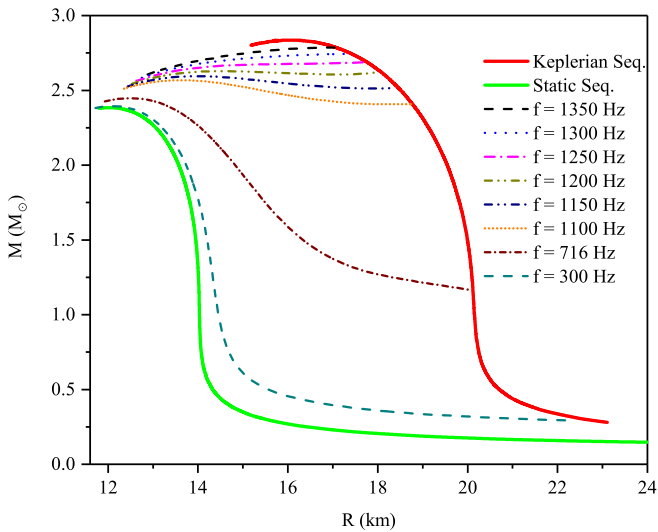}
    \caption{Mass versus equatorial radius along constant high-frequency sequences for the GM1 EOS. It can be seen how above some frequency value (here $\approx 1200$~Hz) the mass at which the constant frequency sequence cuts the Keplerian sequence does not represent a minimum mass bound.}
    \label{fig:GM1}
\end{figure}

One of the most important sequences limiting the equilibrium states of the rotating star is the mass-shedding or Keplerian sequence. This limiting configuration is usually determined by obtaining the configuration for which the velocity of a fluid element at the stellar equator equals the one of a test particle in a stable circular orbit at the equator of the star, namely the limit when the centrifugal and gravitational forces are in balance (see \cite{2003LRR.....6....3S} for a review on rotating stars). At the equator of the star, fluid elements are dislodged at velocities higher than this and the star begins to shed mass. 
Assuming the observed rotation frequency $f$ of the neutron star is not close to the absolute maximum rotation frequency, or that the mass of the star is not near the maximum stable mass (see e.g.~FIG.~\ref{fig:GM1}), then the Keplerian frequency can be assumed as the maximum frequency at which the star can rotate and therefore the condition $f \leq f_K$ imposes a lower limit to $M$, the Keplerian mass, $M_{K}$ or, equivalently, an upper limit to $R$ , the Keplerian radius, $R_{K}$.

\section{Results}

\subsection{EOS-dependent sequence}

\begin{figure} 
	\includegraphics[width=\hsize,clip]{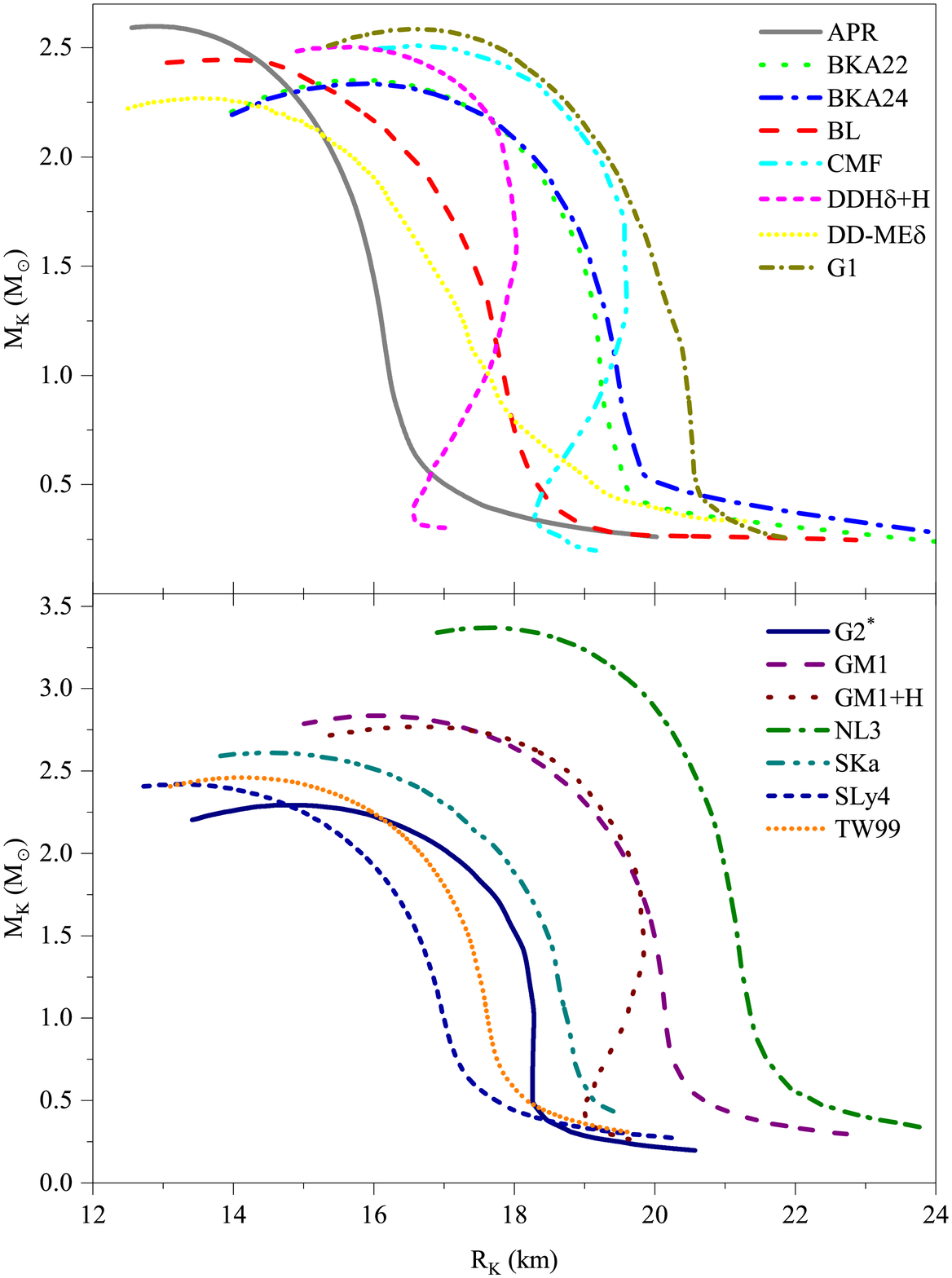}
	\caption{Mass versus equatorial radius of the configurations at the Keplerian sequence for the selected EOS used in this work.}\label{fig:1}
\end{figure}

Despite much studies on high density matter, there is no still agreement on its EOS and a large number of EOS are presented in the literature. Each EOS determines the different equilibrium sequences of static and rotating stars. In FIG.~\ref{fig:1} we show, for the selected EOS, the mass-radius relation of a star in the Keplerian sequence. It is obvious that the sequences depend on the EOS. This apparent strong dependence on the EOS is our motivation to unveil possible universal relations.

\subsection{Universal (EOS-independent) relations}

\begin{figure} 
	\includegraphics[width=\hsize,clip]{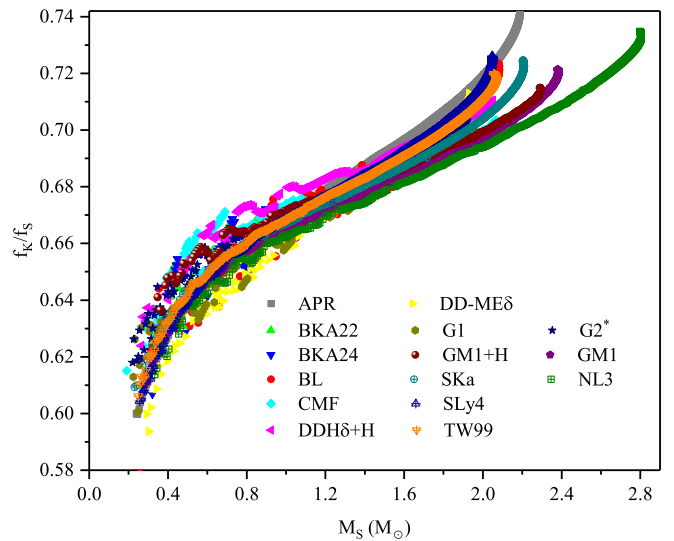}
	\caption{$f_{K}/f_{S}$ ratio versus $M_{S}$ for each selected EOS.}\label{fig:2}
\end{figure}

Lattimer and Prakash (hereafter L\&P) derived in \cite{Lattimer536} a relation, nearly independent of the EOS, which gives the Keplerian frequency of a rotating neutron star, in terms of the radius $R_{S}$ and mass $M_{S}$ of the reference non-rotating neutron star (star with the same central density as the rotating one), providing it is not close to the maximum stable mass allowed by the EOS. The relation is
\begin{equation}
f_{K}=1045\left(\frac{M_{S}}{M_{\odot}}\right)^{1/2}\left(\frac{10 \,{\rm km}}{R_{S}}\right)^{3/2}~{\rm Hz}\approx 0.5701 f_{S},\label{1}
\end{equation}
where $f_{S}=1833(M_{S}/M_{\odot})^{1/2}(10\,{\rm km}/R_{S})^{3/2}$~Hz is the orbital frequency of a test particle spins around a spherical mass $M_{S}$ at a distance $R_{S}$. First, in FIG.~\ref{fig:2} we plot $f_{K}/f_{S}$ ratio against $M_{S}$ and it shows that Eq.~(\ref{1}) underestimates this ratio and the relative error between this relation and the calculated ratio from EOS reaches up to $30\%$ by increasing $M_{S}$. 

\begin{figure} 
	\includegraphics[width=\hsize,clip]{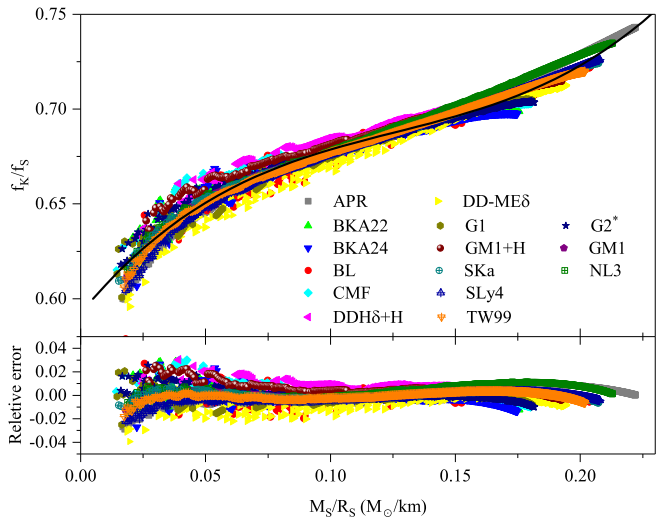}
	\caption{The fitting curve given by Eq. (\ref{freq.}) (black solid curve) and $f_{K}/f_{S}$ ratio versus $M_{S}/R_{S}$ for each selected EOS (upper panel). The relative error between the calculated data and Eq. (\ref{freq.})(lower panel)}\label{fig:3}
\end{figure}

In order to find a more accurate relation for $f_{K}$, we consider as parameter the compactness of the non-rotating configuration of the same central density, $M_{S}/R_{S}$ (dimensionless parameter in geometric units). We plot $f_{K}/f_{S}$ as a function of $M_{S}/R_{S}$ in FIG.~\ref{fig:3}. It becomes clear that this is an approximately EOS-independent and follows the universal relation fitted by
\begin{equation}
y=a_{0}+a_{1}x+a_{2}x^{2}+a_{3}x^{3},
\label{freq.}
\end{equation}
where $y=f_{K}/f_{S}$ and $x=(M_{S}/M_{\odot})({\rm km}/R_{S})$ and the fitting parameters are $a_{0}=0.5926$, $a_{1}=1.5933$, $a_{2}=-9.9582$ and $a_{3}=26.2608$. The order of the polynomial is chosen in such a way that the relative error is small (e.g.~$\lesssim 5\%$) and that the addition of more terms does not improve the fit. In the upper panel of FIG.~\ref{fig:3}, the fitting relation is indicated by a black solid line and, in the lower panel, the relative error between the numerical calculations $y_{cal}$ and the fitting relation $y$, $(y_{cal}-y)/y$ is shown. We conclude the numerical data can be fitted by Eq.~(\ref{freq.}) with a relative error about $2\%$ for a large range of $M_{S}/R_{S}$ and increases up to $4\%$ for $M_{S}/R_{S}\leq 0.05~M_{\odot}/{\rm km}$ which related to $f_{K}\leq 350$~Hz. It is worth to mention that, for some EOS (see e.g. the BL one), the code numerical results show some ``fluctuations'' which could in principle a affect the accuracy of our results. The problem persists even if we use an EOS table with 500 points or more. Fortunately, those fluctuations are observed only in the low mass range, e.g. $M < 1.0~ M_\odot$, which is of no astrophysical relevance. In addition, we have checked that the order of magnitude of such fluctuations in that low-mass region is of the order of $0.1$~km for the equatorial radius. The relative error introduced is therefore $\sim 0.1/20= 0.005$. In the case of the frequency ratio $f_K/f_S$, the fluctuations are of the order of $0.02$, then the relative error introduced is $\sim0.02/0.66=0.03$ which is due to the Keplerian frequency error. This shows that these fluctuations, very likely, are not affecting our numerical results. This is further strengthened by the fact that the majority of EOS of our sample are well-behaved. In the calculations, we have used around 750 points for each EOS, when using more than these number of points does not improve the fitting results. 
.
\begin{figure} 
	\includegraphics[width=\hsize,clip]{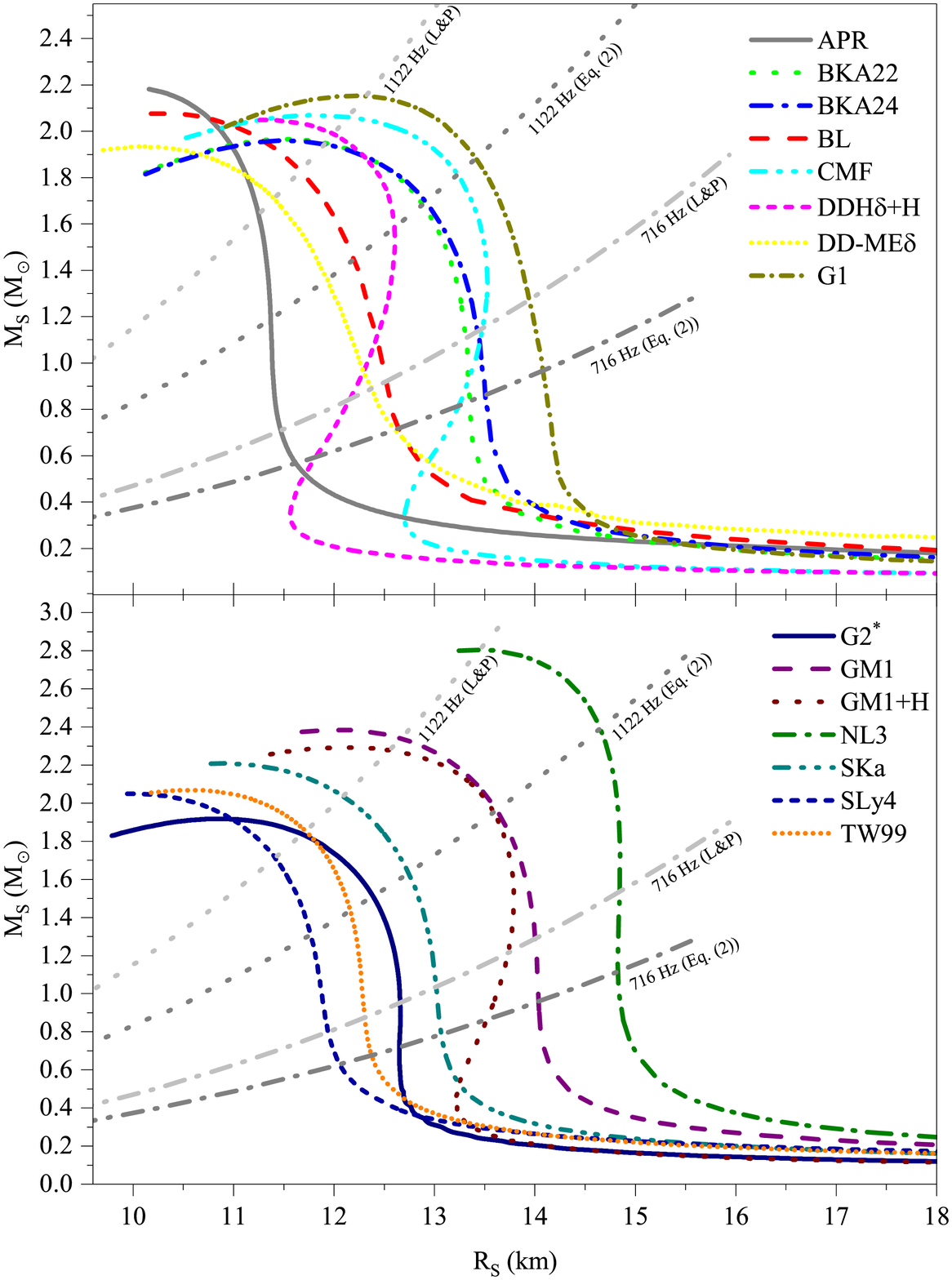}
	\caption{Comparison of the new relation, Eq.~(\ref{freq.}), and L\&P relation, Eq.~(\ref{1}). We here use the fastest observed pulsar up to know, PSR J1748-2446ad \cite{2006Sci...311.1901H} with $f=716$~Hz and a frequency of $f=1122$~Hz, the rotation rate claimed of the neutron star in the XTE J1739-285 \cite{1538-4357-657-2-L97}, but not yet confirmed. The light gray dot-dashed and dotted curves give the relation (\ref{1}) from Ref.~ \cite{Lattimer536}. The dark gray dot-dashed and dotted curves are obtained from our Eq.~(\ref{freq.}). We also show the non-rotating mass-radius relation obtained with all the EOS used in this work.}\label{fig:8}
\end{figure}

We can now compare the new relation, Eq.~(\ref{freq.}), and L\&P relation, Eq.~(\ref{1}). To this aim we use the up-to-know considered to be the fastest known pulsar, PSR J1748--2446ad \cite{2006Sci...311.1901H}, which rotates with a frequency of 716~Hz. In addition to this, for completeness purpose, we do the same analysis using the XTE J1739-285 \cite{1538-4357-657-2-L97}, which has been claimed to rotate with a frequency of 1122~Hz. If these latter observations will be fully confirmed, this pulsar will become the fastest observed neutron star. It can be seen that the approximate relation of L\&P is more stringent than ours, namely the lower limit to $M_S$ it imposes for PSR J1748--2446ad ($f=716$~Hz) is larger than the one set by our Eq. (\ref{freq.}) or, equivalently the upper limit to $R_S$ is lower than ours.

\begin{figure} 
	\includegraphics[width=\hsize,clip]{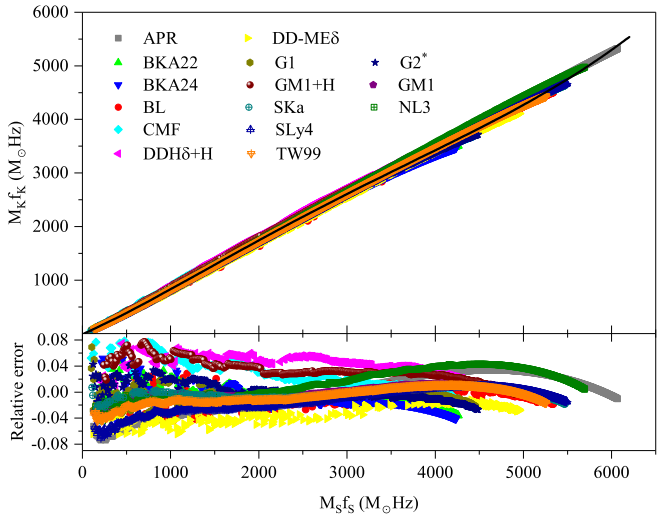}
	\caption{$M_{K}f_{K}$ versus $M_{S}f_{S}$ for each selected EOS. The black solid curve indicates Eq.~(\ref{Eq.Mk/Ms}) (upper panel). The relative error between the curve and data (lower panel).}\label{fig:4}
\end{figure}

\begin{figure} 
	\includegraphics[width=\hsize,clip]{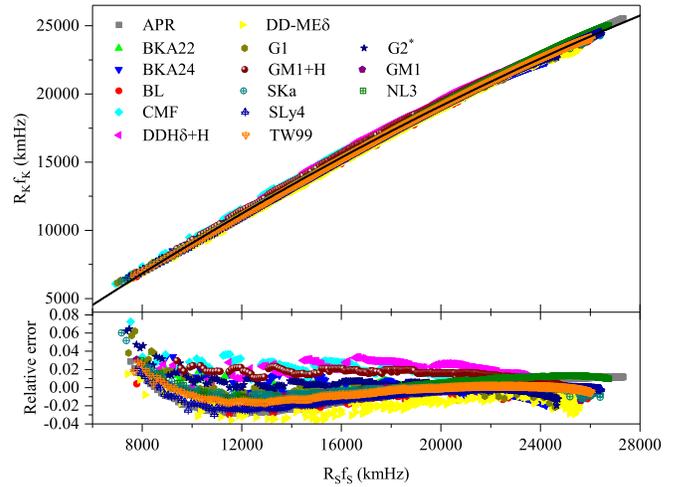}
	\caption{$R_{K}f_{K}$ versus $R_{S}f_{S}$ for each selected EOS. The black solid curve indicates Eq.~(\ref{Eq.Rk/Rs}) (upper panel). The relative error between the curve and data (lower panel).}\label{fig:5}
\end{figure}

This EOS-independent property motivates us to seek for additional independent quantities. We investigate other dimensionless parameters: $M_{K}f_{K}$, $R_{K}f_{K}$ and $M_{K}/R_{K}$ where $M_{K}$ and $R_{K}$ are the mass and circumferential equatorial radius of star in the Keplerian sequence, respectively. FIG.~\ref{fig:4} and \ref{fig:5} show $M_{K}f_{K}$ and $R_{K}f_{K}$ in terms of $M_{S}f_{S}$ and $R_{S}f_{S}$, respectively. It is clear they are approximately EOS-independent. $M_{K}f_{K}$ can be fitted by
\begin{equation}
y=b_{0}+b_{1}x+b_{2}x^{2}+b_{3}x^{3}+b_{4}x^{4}+b_{5}x^{5}
\label{Eq.Mk/Ms}
\end{equation} 
where $y=M_{K}f_{K}$ and $x=M_{S}f_{S}$ are in ${\rm M_{\odot}Hz}$ unit. This equation is shown by a solid curve in the upper panel of FIG.~\ref{fig:4}. The fitting parameters are $b_{0}=-11.4297$, $b_{1}=7.3839\times 10^{-1}$, $b_{2}=1.4973\times 10^{-4}$, $b_{3}=-5.2280\times 10^{-8}$, $b_{4}=6.3072\times 10^{-12}$, $b_{5}=-1.7919\times 10^{-16}$. The relative error between the calculated data and the fitting relation is presented in the lower panel of FIG.~\ref{fig:4}. We can see the data can be fitted by this equation with relative error to about $6\%$.

Another interesting parameter we study is $R_{K}f_{K}$. As FIG.~\ref{fig:5} shows, we fit this data with a relation (black solid curve) which written
\begin{equation}
y=c_{0}+c_{1}x+c_{2}x^{2}+c_{3}x^{3}
\label{Eq.Rk/Rs}
\end{equation} 
where $y=R_{K}f_{K}$ and $x=R_{S}f_{S}$ are in ${\rm kmHz}$ unit. The fitting parameters are $c_{0}=-2.8321\times 10^{3}$, $c_{1}=1.2792$, $c_{2}=-8.6628\times 10^{-6}$ and $c_{3}=-2.0203\times 10^{-11}$. The lower panel of FIG.~\ref{fig:5} shows the relative error between the numerical data and the fit is about $4\%$ for a large range of $R_{S}f_{S}$ and increases up to $8\%$ for $R_{S}f_{S}\leq~8000~{\rm kmHz}$ which related to $f_{K}\leq~350$~Hz and $M_{S}\leq~0.3~M_{\odot}$.

\begin{figure} 
 	\includegraphics[width=\hsize,clip]{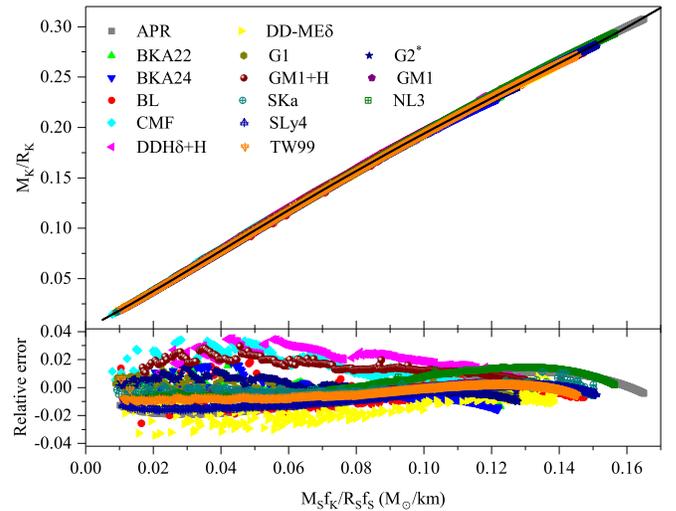}
 	\caption{The dimensionless parameter, compactness, versus $(M_{S}/R_{S})(f_{K}/f_{S})$ for each selected EOS. The black solid curve indicates Eq.~(\ref{Eq.Mk/Rk}) (upper panel). The relative error between the curve and data (lower panel).}\label{fig:6}
 \end{figure}

The next parameter we study is the compactness along the Keplerian sequence, $M_{K}/R_{K}$. FIG.~\ref{fig:6} shows in the upper panel the numerical data and the fitting relation (block solid curve). The relative error, which is around $4\%$, is shown in the lower panel. 

The numerical relation is well fitted by
\begin{equation}
y=d_{1}x+d_{2}x^{2}+d_{3}x^{3}+d_{4}x^{4}
\label{Eq.Mk/Rk}
\end{equation} 
where $x=(M_{S}/M_\odot)({\rm km}/R_{S})(f_{K}/f_{S})$ is a dimensionless quantity and the fitting parameters are obtained in such a way that $y=M_{K}/R_{K}$ is also dimensionless. The fitting parameters are $d_{1}=1.7658$, $d_{2}=7.0424$, $d_{3}=-7.5442\times 10^{1}$ and $d_{4}=2.2236\times 10^{2}$. This equation, which shows the relation between the compactness of rotating neutron star in the Keplerian sequence and that of the reference static neutron star, gives the least compact rotating neutron star for a given rotation frequency.

\begin{figure} 
	\includegraphics[width=\hsize,clip]{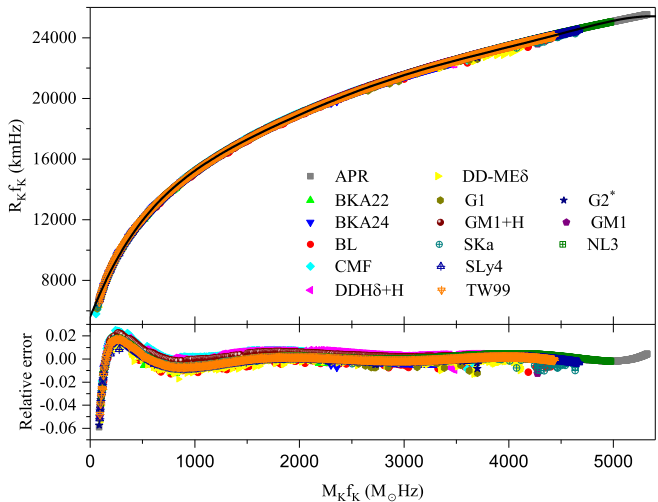}
	\caption{$R_{K}f_{K}$ versus $M_{K}f_{K}$ for each selected EOS. The black solid curve indicates Eq.~(\ref{Eq.Rkfk}) (upper panel). The relative error between the curve and data (lower panel).}\label{fig:7}
\end{figure}
 
The introduced equations give the relation between the various quantities of rotating and static neutron stars. The following equations give the relation between dimensionless parameters, e.g. $R_{K}f_{K}$ and $M_{K}f_{K}$. FIG.~\ref{fig:7} shows that these quantities can be related in an EOS-independent fashion and a fitting relation is
\begin{equation}
y=e_{0}+e_{1}x+e_{2}x^{2}+e_{3}x^{3}+e_{4}x^{4}+e_{5}x^{5}+e_{6}x^{6},
\label{Eq.Rkfk}
\end{equation}
where $y=R_{K}f_{K}$, $x=M_{K}f_{K}$ are in ${\rm kmHz}$ and ${\rm M_{\odot}Hz}$ units, respectively. The fitting parameters are $e_{0}=56.8408\times 10^{2}$, $e_{1}=17.5480$, $e_{2}=-1.2587\times 10^{-2}$, $e_{3}=5.9763\times 10^{-6}$, $e_{4}=-1.5983\times 10^{-9}$, $e_{5}=2.1962\times 10^{-13}$ and $e_{6}=-1.2032\times 10^{-17}$. The relative error is about $1\%$ for $M_{K}f_{K}\geq 500~M_{\odot}\,{\rm Hz}$ and increases to $4\%$ for smaller amounts. We can also reverse the above relation to obtain
\begin{equation}
y=h_{0}+h_{1}x+h_{2}x^{2}+h_{3}x^{3}+h_{4}x^{4}+h_{5}x^{5}+h_{6}x^{6},
\label{Eq.Mkfk}
\end{equation}
where $y=M_{K}f_{K}$ and $x=R_{K}f_{K}$. The fitting parameters are $h_{0}=-28.1955\times 10^{2}$, $h_{1}=1.3433$, $h_{2}=-2.5528\times 10^{-4}$, $h_{3}=2.5133\times 10^{-8}$, $h_{4}=-1.3087\times 10^{-12}$, $h_{5}=3.5361\times 10^{-17}$ and $h_{6}=-3.8115\times 10^{-22}$. This relation fit the numerical data to about $2\%$ for $R_{K}f_{K}\geq 1000$~km~Hz and the relative error increases to $13\%$ for smaller $R_{K}f_{K}$.  

    \label{fig:GM1sequences}
\section{Conclusions}

We studied the Keplerian (mass-shedding) sequence of uniformly rotating neutron stars by using a variety of EOS. We searched for nearly EOS-independent relations that connect structure properties of the star to each other.

We showed that the relation (\ref{1}) by L\&P \cite{Lattimer536}, namely the ratio between the rotation frequency of a neutron star at the Keplerian sequence, $f_K$, and the Keplerian frequency $f_S$ of a particle in a circular orbit at the surface of the non-rotating neutron star with the same central density as the rotating one, is neither accurate nor EOS-independent as originally thought at the time it was proposed (see FIG.~\ref{fig:3}). We thus found the new relation given by Eq.~(\ref{freq.}) of the ratio $f_K/f_S$ as a function of the non-rotating neutron star compactness, $M_S/R_S$, which is EOS-independent within only a $4\%$ error. We use this new relation to put new constraints to the neutron star mass-radius relation using the fastest pulsar observed (see FIG.~\ref{fig:8}).

We proceeded to search for additional (nearly) EOS-independent relations and derive fitting polynomials for $M_{K}f_{K}$ in terms of $M_{S}f_{S}$ (see Eq.~\ref{Eq.Mk/Ms} and FIG.~\ref{fig:4}); $R_{K}f_{K}$ in terms of $R_{S}f_{S}$ (see Eq.~\ref{Eq.Rk/Rs} and FIG.~\ref{fig:5}); for $M_K/R_K$ in terms of $M_S f_{K}/(R_S f_S)$ (see Eq.~\ref{Eq.Mk/Rk} and FIG.~\ref{fig:6}); and for $R_K f_K$ in terms of $M_K f_K$ (see Eq.~\ref{Eq.Rkfk} or \ref{Eq.Mkfk} and FIG.~\ref{fig:7}). These new fitting relations are nearly EOS-independent within a maximum error of $8\%$.

The universality of the Keplerian sequence properties found in this article, in particular Eqs.~(\ref{freq.}), (\ref{Eq.Mk/Rk}) and (\ref{Eq.Rkfk} or Eq.~\ref{Eq.Mkfk}) which have been shown to be most accurate, add to the set of other universal relations of neutron stars known in the literature such as the $I$-Love-$Q$ relation \cite{PhysRevD.88.023009,Yagi365}, the binding energy of non-rotating and rotating neutron stars \cite{2015PhRvD..92b3007C} and the nearly EOS-independent behavior of the energy, angular momentum and radius of the last circular orbit of a test-particle around rotating neutron stars  \cite{PhysRevD.96.024046,0004-637X-861-2-141}. This set of universal, analytic formulas, facilitate the inclusion of general relativistic effects in the description of relativistic astrophysical systems involving fast rotating neutron stars (see e.g.~\cite{2015ApJ...812..100B,2016ApJ...833..107B,2018arXiv180304356B}) and can be also used to put constraints to the mass-radius relation of neutron stars and so to the EOS of nuclear matter (see FIG.~\ref{fig:8}).


\section{Acknowledgments}
R.R thanks ICRANet headquarters in Pescara for hospitality and the Faculty members and researchers for valuable discussions. We thank the team members of LORENE for developing the public code.

\bibliographystyle{apsrev}
\bibliographystyle{apsrev}

\end{document}